\documentclass[twocolumn,secnumarabic,amssymb, amsmath, nobibnotes, aps, pra, nofootinbib]{revtex4-2}
\usepackage{graphics}      % standard graphics specifications
\usepackage{graphicx}      % alternative graphics specifications
\usepackage{longtable}     % helps with long table options
\usepackage{bm}            % special 'bold-math' package
\usepackage{natbib}
\usepackage{physics}
\usepackage{xcolor}
\usepackage[caption=false]{subfig}
\usepackage{dcolumn}
\usepackage{hyperref}
\usepackage{amsmath}
\usepackage{amssymb}
\usepackage{slashed}
\usepackage{mathtools}
\usepackage[thinc]{esdiff}

\begin{document}

\title{Vacuum polarization correction to atomic energy levels in the path integral formalism}%

\author{Sreya Banerjee}%
\email{banerjee@mpi-hd.mpg.de}
\affiliation{Max Planck Institute for Nuclear Physics, Heidelberg, Germany}
\author{Z. Harman}
\email{harman@mpi-hd.mpg.de}
\affiliation{Max Planck Institute for Nuclear Physics, Heidelberg, Germany}
\date{\today}%

\begin{abstract}
Vacuum polarization corrections to the energy levels of bound electrons are calculated using a perturbative path integral formalism. We apply quantum electrodynamics
in a framework which treats the strong binding nuclear field to all orders.
The effective potential, derived from the Dyson-Schwinger equation for the photon propagator, is then considered pertubatively.
Expressions for the vacuum polarization shift of binding energies is obtained from the poles of the spectral function up to
second order. Numerical results are provided to select candidates for novel tests of strong-field quantum electrodynamics by means of precision mass spectrometry.
\end{abstract}

\maketitle

\section{Introduction}

Radiative corrections to energy levels of atoms, and, more recently, highly charged ions (HCI), have been at the focal point of theoretical and experimental
studies over the past decades. At the one-loop level, there are two types of contributions, the vacuum polarization (VP) and the self-energy effect, which both
scale, to leading order, with $Z^4$, i.e. with the forth power of the atomic number $Z$, rendering heavy HCI an ideal test bench for the study of radiative effects.
Experimental advances in the production, trapping and storing of HCI and the investigation of their properties with unprecedented accuracy
\cite{Gumberidze2005,Sturm2011,PhysRevLett.85.3109,Micke2020,Kubicek2014,Sailer_2022,Gillaspy2010,Gillaspy1996,Marmar1986,Deslattes1984,
Beiersdorfer1989,Decaux1997,Tavernier1985} call for versatile theoretical frameworks for the description of such systems. Very recently, precision mass spectrometry
with Penning-trap setups have reached the 1-eV level of accuracy~\cite{Rischka2020,Schuessler2020,Filianin2021,Kromer2022}, allowing the study of QED effects in heavy HCI by
means of electron binding energy determinations through mass measurements and the mass-energy equivalence relation.

In this article, we develop a functional integral formalism for the calculation of VP corrections for bound atomic states in the Furry picture~\cite{PhysRev.81.115}. 
The usual perturbative framework of quantum field theory overlooks the intricacies of the non-perturbative effects that show up in the study of bound systems. We overcome
this existing disagreement between the intrinsically nonpertubative bound states and the perturbative nature of quantum electrodynamics (QED). We use Feynman's path
integral formalism~\cite{Feynman:100771} in the relativistic regime
\cite{PhysRev.80.440, PhysRev.84.108, PhysRevD.49.4049, PhysRevD.34.2323, PhysRevD.44.3230, Ferraro:1992eb, PhysRevD.38.2468, PhysRevD.48.748}, wherein, the time-sliced
formulation of path integrals ensures Lorentz invariance. We treat the path integrals perturbatively and we
proceed by summing the perturbative expansion to all orders~\cite{BHAGWAT1989417,Schulman1981TechniquesAA}.

The introduction of functional integral methods to atomic systems is also motivated by the prospects of incorporating non-electromagnetic interactions
into the precision theory of atoms and ions. As an example, hadronic vacuum polarization effects have also been computed by means of an ab initio quantum chronodynamic (QCD)
Schwinger-Dyson approach~\cite{Goecke2011,Bashir2012}. The continuing increase in experimental accuracy may necessitate in future the inclusion of such QCD corrections in atomic
spectra~\cite{Friar1999,Karshenboim2021,Breidenbach2022,Dizer2023}. Furthermore, prospects of new physics searches with low-energy atomic precision experiments
(see e.g.~\cite{Jaeckel2010,Flambaum2018,Berengut2018,Counts2020,Sailer_2022}) also suggest to employ a versatile field theoretical formalism enabling the inclusion of various
types of gauge-boson propagators.

This article is organized as follows. In Section~\ref{sec:vacpo}, we derive the effective potential describing the VP correction to the nuclear potential in the framework
of path integrals using the Schwinger-Dyson equations. In Section~\ref{sec:formulation}, we discuss the general perturbative formulation of path integrals.
In Section~\ref{sec:uehling}, we describe our computations of the
VP contributions to the Lamb shift of energy levels using the formalism introduced in Section~\ref{sec:formulation}. In Section~\ref{sec:results},
we tabulate and discuss our numerical results of the Uehling contribution to the energy shift and provide concluding remarks.
Through the article, we use natural units, unless stated otherwise.

\section{Schwinger-Dyson equations for the photon self-energy}
\label{sec:vacpo}
We begin by summarizing the derivation of the complete expression for the VP correction to the photon propagator, also known as the photon self-energy. This is performed by defining
the Schwinger-Dyson equation for the photon propagator using path integrals in analogy to~\cite{PhysRevD.83.045007}.
The gauge-fixed Lagrangian of quantum electrodynamics (QED) is given as
\begin{eqnarray}
	\mathcal{L}_{\text{QED}}(x)=&& A_{\mu}(x)\left[g^{\mu\nu}\partial^2+\left(\frac{1}{\xi}-1\right)\partial^\mu\partial^\nu\right]A_{\nu}(x)\nonumber\\
	&&\qquad+\bar{\psi}(x)(i\slashed{D}-m)\psi(x)\,,
	\label{eq:1}
\end{eqnarray}
where $A_\mu$ is the gauge field operator for the photon field, $\psi(x)$ is the Dirac field of the electron (or other bound lepton) in coordinate space, $m$ is its bare mass,
$g^{\mu\nu}$ is the metric tensor with the $\mu,\nu \in \left\{0,1,2,3\right\}$ being Lorentz indices, $\slashed{D}=\partial_\mu-ieA_\mu$, and $\xi$ the gauge fixing parameter.

The generating functional $Z$ is constructed using this Lagrangian, and the external source $J_\mu$ of the photon field, and the sources $\eta$ and
$\bar{\eta}$ of the fermion fields $\bar{\psi}$ and $\psi$, respectively, as the functional integral 
\begin{eqnarray}
	Z[\eta,\bar{\eta},J_\mu]&&=\int\mathcal{D}\bar{\psi}\mathcal{D}\psi\mathcal{D}A \exp \biggl\{i\int d^4 x[\mathcal{L}_{\text{QED}}\label{eq:2}\\
	&&+J_\mu(x) A^\mu(x)+\bar{\psi}(x)\eta(x)+\bar{\eta}(x)\psi(x)]\biggr\}\,.\nonumber
\end{eqnarray}
The fermion fields and their sources are anti-commuting Grassmann variables.
To arrive at the Schwinger-Dyson equation, we consider that the functional integral of a total derivative is zero,
\begin{eqnarray}
	\int \mathcal{D}[\phi]\fdv{\phi}=0\,,
\end{eqnarray}
where $\phi$ is an arbitrary field variable.
For the photon propagator, the derivative is taken with respect to the gauge field $A_\mu(x)$:
\begin{eqnarray}
	&&\int\mathcal{D}[\bar{\psi}\psi A]\frac{\delta}{\delta A_\mu(x)}\exp\{i[S(\bar{\psi},\psi, A)\label{eq:3}\\
	&&+\int d^4 x J_\mu(x) A^\mu(x)+\bar{\psi}(x)\eta(x)+\bar{\eta}(x)\psi(x)]\}=0\,.\nonumber
\end{eqnarray}
Eq.~(\ref{eq:3}) can be written as a differential equation for the generating functional $Z$:
\begin{eqnarray}
	\Biggl[\frac{\delta S}{\delta A_\mu(x)}&&\left(-i\frac{\delta}{\delta J_\mu }, i\frac{\delta}{\delta \eta}, -i\frac{\delta}{\delta \bar{\eta}}\right)\label{eq:4}\\
	&&\qquad\qquad\qquad+J^\mu(x)\Biggr]Z[\eta,\bar{\eta},J_\mu]=0\,,\nonumber
\end{eqnarray}
where we have established a correspondence between fields and their source terms through functional derivatives
\begin{eqnarray}
	\psi(x)  \leftrightarrow -\frac{i\delta}{\delta \bar{\eta}(x)}\,,\quad \bar{\psi}(x)\leftrightarrow \frac{i\delta}{\delta {\eta}(x)}\,,\quad
        A^\mu(x) \leftrightarrow -\frac{i\delta}{\delta J_\mu(x)}\,. \nonumber
\end{eqnarray}
The first term on the left-hand side of Eq.~(\ref{eq:4}) is solved by implementing the G\^ateux derivative method:
\begin{eqnarray}
	\dv{S(\phi+\epsilon\tau)}{\epsilon}|_{\epsilon=0}^{}=\int\dd[4]{x}\frac{\partial S}{\partial \phi} \tau\,\,\,,
\end{eqnarray}
yielding
\begin{eqnarray}
	&&\frac{\delta S}{\delta A_\mu(x)}\label{eq:6}\\
	&&=\left\{\left[g^{\mu\nu}\partial^2+\left(\frac{1}{\xi}-1\right)\partial^\mu\partial^\nu\right]A_\nu(x)+e\psi(x)\gamma^\mu \bar{\psi}(x)\right\}\,.\nonumber
\end{eqnarray}
Thus, Eq.~(\ref{eq:4}), written in terms of the source terms, becomes
\begin{eqnarray}
	&&\biggl\{\left[g^{\mu\nu}\partial^2+\left(\frac{1}{\xi}-1\right)\partial^\mu\partial^\nu\right]\left(\frac{-i\delta}{\delta J^\nu(x)}\right)\\
	&&\qquad\quad+e\frac{-i\delta}{\delta \bar{\eta}(x)}\gamma^\mu \frac{i\delta}{\delta {\eta}(x)}+J^\mu(x)\biggr\}Z[\eta,\bar{\eta},J_\mu]=0\,.\nonumber
	\label{eqZ}
\end{eqnarray}
In terms of the generating functional $W$ for the Green's functions of the connected Feynman diagrams,  $W[\eta,\bar{\eta},J_\mu]= -i\ln Z[\eta,\bar{\eta},J_\mu]$,
Eq.~(\ref{eqZ}) has the form
\begin{eqnarray}
	&&\biggl\{\left[g^{\mu\nu}\partial^2+\left(\frac{1}{\xi}-1\right)\partial^\mu\partial^\nu\right]\left(\frac{\delta}{\delta J^\nu(x)}\right)\label{eq:8}\\
	&&+ie\frac{-i\delta}{\delta \bar{\eta}(x)}\gamma^\mu \frac{i\delta}{\delta {\eta}(x)}\biggr\}W-ie\frac{-i\delta W}{\delta \bar{\eta}(x)}\gamma^\mu \frac{i\delta W}{\delta {\eta}(x)}=-J^{\mu}(x)\,.\nonumber
\end{eqnarray}
We now introduce the effective action $\Gamma$, defining the one-particle-irreducible (1PI) Green's functions, in terms of the generating functional $W$ through a
Legendre transformation, following~\cite{PhysRevD.83.045007}:
\begin{equation}
	\Gamma[\bar{\psi}, \psi, A_\mu]=W[\eta, \bar{\eta}, J_\mu]-\int\dd[4]{x}(\bar{\psi}\eta+\psi\bar{\eta}+A^{\mu}J_\mu)\,.
	\label{eq:9}
\end{equation}
Owing to the above transformation, we can define the field terms $(\bar{\psi}, \psi, A_\mu)$ in terms of the source terms $(\eta, \bar{\eta}, J_\mu)$:
\begin{eqnarray}
	&&A_\mu=-i\frac{\delta W[J]}{\partial J^{\mu}}\,,\quad \psi=-i\frac{\delta W[\eta]}{\delta \bar{\eta}}\,,\quad\bar{\psi}=i\frac{\delta W[\bar{\eta}]}{\delta \eta}\,.\nonumber\\\nonumber\\
	&&J_\mu=-\frac{\delta \Gamma[A]}{\partial A^{\mu}}\,,\quad \eta=-\frac{\delta \Gamma[\bar{\psi}]}{\delta {\psi}}\,,\quad\bar{\eta}=\frac{\delta \Gamma[{\psi}]}{\delta \bar{\psi}}\,.\nonumber
\end{eqnarray}
Setting the fermion sources to zero and using the above expressions, Eq.~(\ref{eq:8}) becomes
\begin{eqnarray}
	&&\frac{\delta \Gamma[A]}{\partial A_{\mu}(x)}=i\left[g^{\mu\nu}\partial^2+\left(\frac{1}{\xi}-1\right)\partial^\mu\partial^\nu\right]A_\mu(x)\nonumber\\
	&&\qquad\qquad\qquad\qquad\qquad+ie\frac{-i\delta}{\delta \bar{\eta}(x)}\gamma^\mu \frac{i\delta}{\delta {\eta}(x)}W\,.
	\label{eq:10}
\end{eqnarray}
From the above expression, the connected two-point function -- or, in this case, the complete electron propagator -- in an external field $A_\mu$ can be identified easily as
\begin{equation}
	S(x,y)= i\frac{\delta^2 W[\eta,\bar{\eta},J_\mu]}{\delta \eta(y)\delta \bar{\eta}(x)}|_{\psi=\bar{\psi}=0}=\left(\frac{\delta^2 \Gamma}{\delta \psi(x)\delta \bar{\psi}(y)}\right)^{-1}\,.
	\label{eq:11}
\end{equation}
Using the identity as derived in Ref.~\cite{Faizullaev_1995},
\begin{eqnarray}
	\frac{-i\delta}{\delta \bar{\eta}(x)}\gamma^\mu \frac{i\delta W}{\delta {\eta}(y)}&&=\Tr{\frac{-i\delta}{\delta \bar{\eta}(x)}\gamma^\mu \frac{i\delta}{\delta {\eta}(y)}W[\eta, \bar{\eta}, J^\mu]}\nonumber\\\nonumber\\
	&&=-\Tr{\gamma^\mu S(x,y)}\,,
	\label{eq:12}
\end{eqnarray}
(with $\Tr{\dots}$ representing the trace of the matrix in the curly brackets) and Eq.~(\ref{eq:11}), Eq.~(\ref{eq:10}) reduces to
\begin{eqnarray}
	&&\frac{\delta \Gamma[A]}{\partial A_{\mu}(x)}\label{eq:13}\\
	&&=i\left[g^{\mu\nu}\partial^2+\left(\frac{1}{\xi}-1\right)\partial^\mu\partial^\nu\right]A_\mu(x)-ie\Tr{\gamma^\mu S(x,x)}\,.\nonumber
\end{eqnarray}
Since our aim is to determine the photon propagator, we take the second derivative of the effective action $\Gamma$ with respect to the photon field,
and thus Eq.~(\ref{eq:13}) changes to
\begin{eqnarray}
	&&\frac{\delta^2 \Gamma[A]}{\partial A_{\mu}(x)\delta A_\nu(y)}=i\left[g^{\mu\nu}\partial^2+\left(\frac{1}{\xi}-1\right)\partial^\mu\partial^\nu\right]\delta(x-y)\nonumber\\
	&&\qquad\qquad\,-ie\Tr{\gamma^{\mu}\frac{\delta}{\delta A_\nu(y)}\left(\frac{\delta^2 \Gamma}{\delta \psi(x)\delta \bar{\psi}(x)}\right)^{-1}}\,.
	\label{eq:14}
\end{eqnarray}
Here, $\delta (x-y)$ represents a four-dimensional Dirac delta. The derivative of an inverse matrix is given as
\begin{eqnarray}
	\dv{B^{-1}}{t}=-B^{-1}\dv{B}{t}B^{-1}\,,\nonumber
\end{eqnarray}
where 
\begin{eqnarray}
	B=\frac{\delta^2 \Gamma}{\delta \psi(x)\delta \bar{\psi}(x)}\,.\nonumber
\end{eqnarray}
Applying this to the functional derivative in Eq.~(\ref{eq:14}), referring to Eq.~(\ref{eq:11}), and considering that similar to the fermionic case,
\begin{eqnarray}
	\frac{\delta^2 \Gamma[A]}{\partial A_{\mu}(x)\delta A_\nu(y)}=D_{\mu\nu}^{-1}(x-y)\,,\nonumber
\end{eqnarray}
we obtain a modified version of Eq.~(\ref{eq:14}) for the inverse of the photon propagator:
\begin{eqnarray}
	&&D_{\mu\nu}^{-1}(x-y)=i\left[g_{\mu\nu}\partial^2+\left(\frac{1}{\xi}-1\right)\partial_\mu\partial_\nu\right]\delta(x-y)\label{eq:15}\\
	&&-ie\int \dd[4]{x_1}\dd[4]{x_2}\Tr\biggl\{\gamma_\mu\left(\frac{\delta^2 \Gamma}{\delta \bar{\psi}(x)\delta {\psi}(x_1)}\right)^{-1}\nonumber\\
	&&\qquad\cross\left(\frac{\delta^3 \Gamma}{\delta A_\nu(y)\delta \bar{\psi}(x_1)\delta {\psi}(x_2)}\right)\left(\frac{\delta^2 \Gamma}{\delta \psi(x)\delta \bar{\psi}(x_2)}\right)^{-1}\biggr\}\,.\nonumber
\end{eqnarray}
The third derivative of $\Gamma$ gives us the connected three-point Green's function, or the electron-photon vertex function
\begin{eqnarray}
	\left(\frac{\delta^3 \Gamma}{\delta A_\nu(y)\delta \psi(x_1)\delta \bar{\psi}(x_2)}\right)=e\Gamma_\nu(y;x_1,x_2)\,,
	\label{eq:16}
\end{eqnarray}
and using the expressions for the complete electron propagator from Eq.~(\ref{eq:11}), we can write Eq.~(\ref{eq:16}) as
\begin{eqnarray}
	&&D_{\mu\nu}^{-1}(x-y)=i\left[g_{\mu\nu}\partial^2+\left(\frac{1}{\xi}-1\right)\partial_\mu\partial_\nu\right]\delta(x-y)\label{eq:17}\\
	&&\quad-ie^2\int\dd[4]{x_1}\dd[4]{x_2}\Tr\biggl\{S(x_1,x)\gamma_\mu S(x,x_2)\Gamma_\nu(x_2,x_1;x)\biggr\}\,.\nonumber
\end{eqnarray}
The second term on the right-hand side is the VP tensor in coordinate space:
\begin{eqnarray}
	&&\Pi_{\mu \nu}(x,y)\label{eq:18}\\
	&&\equiv -ie^2\int\dd[4]{x_1}\dd[4]{x_2}\Tr\biggl\{S(x_1,x)\gamma_\mu S(x,x_2)\Gamma_\nu(x_2,x_1;x)\biggr\}\,.\nonumber
\end{eqnarray}
Fourier transforming to momentum space, we obtain \cite{Greiner2009} the polarization tensor as a function of the photon four-momentum $k$,
\begin{equation}
	\Pi_{\mu \nu}(k)=-ie^2\int\frac{\dd[4]{p}}{(2\pi)^4}\Tr\biggl\{S(p)\gamma_\mu S(p-k)\Gamma_\nu(p,k;p-k)\biggr\}\,,
	\label{eq:19}
\end{equation}
where $p$ is the four-momentum of one of the virtual leptons. The inverse photon propagator in momentum space thus becomes
\begin{eqnarray}
	D_{\mu\nu}^{-1}(k)=i\left[g_{\mu\nu}k^2+\left(\frac{1}{\xi}-1\right)k_\mu k_\nu\right]+\Pi_{\mu \nu}(k) \,.
	\label{eq:20}
\end{eqnarray}
This gives us the Schwinger-Dyson equation for the unquenched (dressed) photon propagator in momentum space, where the second term on the right-hand side describes, to lowest order,
the photon propagator perturbed by the creation and annihilation of a virtual lepton-antilepton pair.
\begin{figure}
	\centering
	\includegraphics[width=0.48\textwidth]{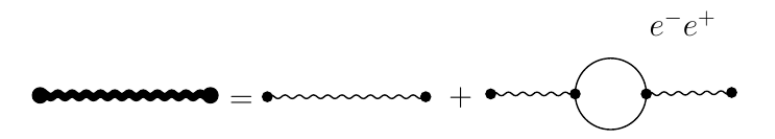}
	\caption{Diagrammatic representation of the Schwinger-Dyson equation for the photon propagator.}
	\label{fig:photonprop}
\end{figure}
Considering that the photon self-energy or the VP is necessarily transverse~\cite{Peskin:1995ev,Breckenridge_1995}, we obtain for the full photon propagator,
as illustrated in Fig.~\ref{fig:photonprop}, the equation
\begin{equation}
	D_{\mu\nu}(k)=-i\left[\frac{g_{\mu\nu}}{k^2}-\frac{k_\mu k_\nu}{k^4}\right]\frac{1}{1+\Pi(k^2)}-i\xi\frac{k_\mu k_\nu}{k^4}\,.
	\label{eq:21}
\end{equation}
The gauge fixing parameter is set to $\xi= 0,1$ in the Landau and Feynman gauges, respectively.
$ \Pi(k^2) $ is the polarization function, defined as in e.g. \cite{Greiner2009} as $ \Pi_{\mu \nu}(k)= (k_\mu k_\nu -g_{\mu\nu} k^2)\Pi(k^2) $.
The well-known free photon propagator can be reproduced ed by setting $\Pi(k^2)=0$.

One can now calculate the four-potential~\cite{Greiner2009,Peskin:1995ev} induced by a charge density $j^{\nu}(q)$,
\begin{eqnarray}
	A_\mu^{'}(x)=\int\frac{\dd[4]{k}}{(2\pi)^4} e^{-ik\cdot x} D_{\mu\nu}(k) j^{\nu}(k)\,,
	\label{eq:23}
\end{eqnarray}
with $D_{\mu\nu}(k)$ being the complete, dressed photon propagator. In our special case of a static nucleus, $j^{\nu}(q) = -Ze\delta_{\nu 0}$.
Taking the VP-perturbed propagator to \textit{leading order}, the interaction potential in Eq.~(\ref{eq:23}) is reduced following \cite{Greiner2009} to the well-known expression
with an integration variable $y$
\begin{eqnarray}
	A^{'}_0(r)&=&-\frac{Ze}{r}\biggl[1+\frac{2\alpha}{3\pi}\int_1^{\infty}dy\left(1+\frac{1}{2y^2}\right)\frac{\sqrt{y^2-1}}{y^2}e^{-2my r}\biggr]\nonumber\\
	&\equiv&-\frac{Ze}{r}+U_{\rm VP}^{\rm Ueh}(r)\,,
	\label{eq:24}
\end{eqnarray}
with $\alpha$ being the fine-structure constant.
This interaction potential is a sum of two distinct terms, with the second defining the Uehling potential $U_{\rm VP}^{\rm Ueh}$.
Following the work of \cite{Frolov2012}, the integral in Eq.~(\ref{eq:24}) is reduced using the modified Bessel functions (or Bickley-Naylor functions), $Ki$,
and defining integrals to the form
\begin{align}
	&I_U(a)=\int_1^{\infty}d\xi\left(1+\frac{1}{2\xi^2}\right)\frac{\sqrt{\xi^2-1}}{\xi^2}e^{-a\xi }\,\label{eq:26}\\
	\quad&=\int_0^{\infty}dx\,e^{-a\cosh(x)}\left(1-\frac{1}{2\cosh^2(x)}-\frac{1}{2\cosh^4(x)}\right)\label{eq:27}\,,\\
	&=Ki_0(a)-\frac{1}{2}Ki_2(a)-\frac{1}{2}Ki_4(a)\label{eq:28}\,,
\end{align}
where $a=2\alpha^{-1}r, \text{and}\,Ki_n(z)=\int_{0}^{\infty}\frac{e^{-z\,\cosh(x)}}{\cosh^n x}dx\,, n \geq 1$.
Using these functions, the closed-form expression of the Uehling potential is obtained as in \cite{Frolov2012}
\begin{align}
	&U_{\rm VP}^{\rm Ueh}(a)=-\frac{2Ze\alpha}{3\pi r} \biggl[\left(1+\frac{a^2}{12}\right)K_0(a)\nonumber\\
	&\qquad\qquad\qquad-\frac{a}{12}Ki_1(a)-\left(\frac{5}{6}+\frac{a^2}{12}\right)Ki_2(a)\biggr]\,.
	\label{eq:29}
\end{align}
We work with this closed expression for the vacuum polarization potential to obtain the corresponding energy shift using path integral formalism,
as outlined in the following Section. We note that higher-order terms in the dressed photon propagator would lead to the K\"all\'en-Sabry type corrections~\cite{Kallen1955} to the potential.

\section{Perturbative path integrals for bound states}
\label{sec:formulation}
\begin{figure}
	\centering
	\includegraphics[width=0.45\textwidth]{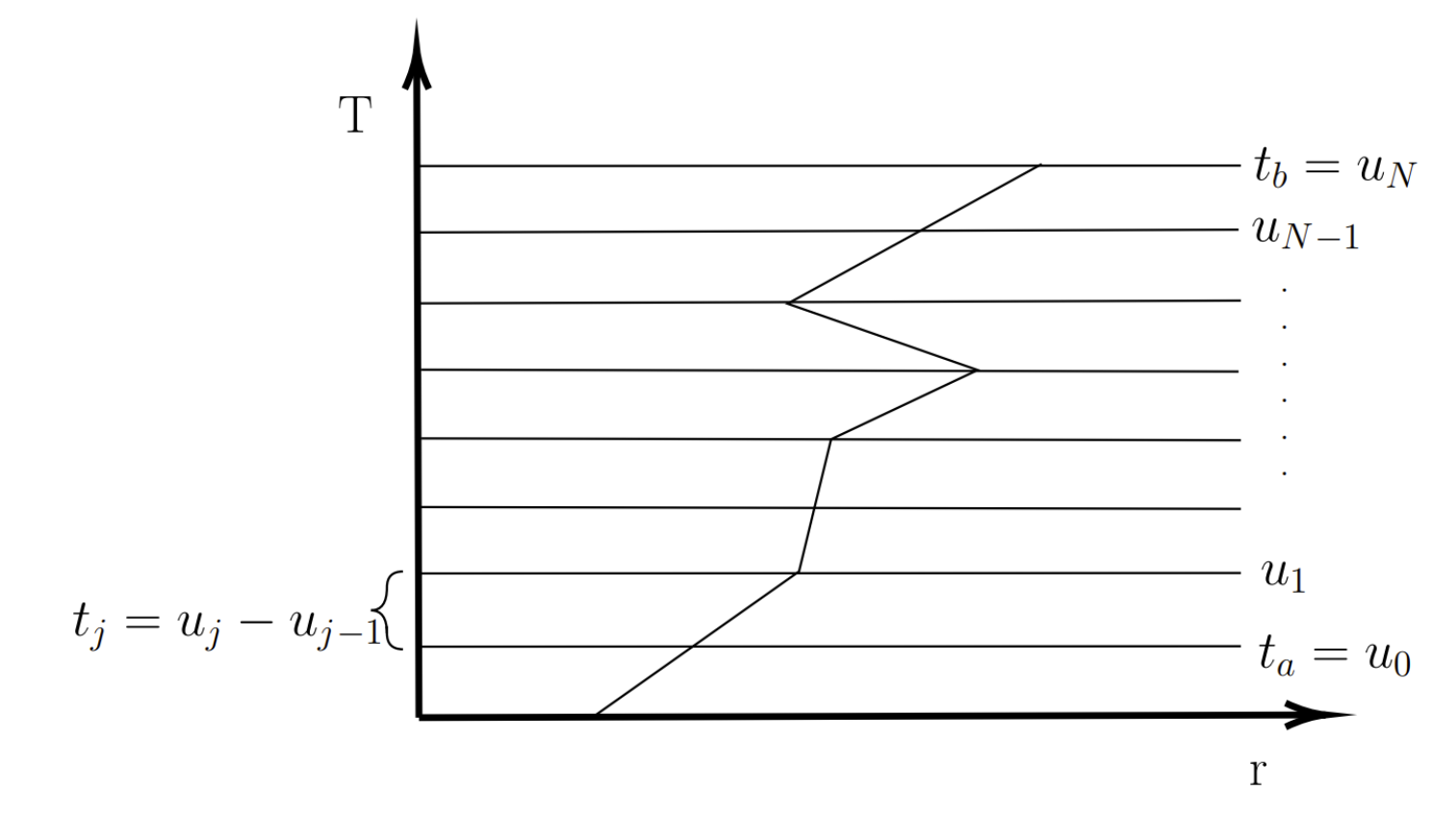}
	\caption{Time-lattice division for a given path. The lattice is divided into $N+1$ equidistant points, with the width given by $t_j$.}
	\label{fig:timelat}
\end{figure}

This Section lays the groundwork for the \emph{perturbative approach} to the path integral formalism, following the articles \cite{Schulman1981TechniquesAA,BHAGWAT1989417, CAMBLONG_2004}. 
We begin with the modification of the relativistic action integral, wherein the action $S$ is perturbed by a potential $\Delta V$, and the perturbation expansion is summed to all orders:
\begin{eqnarray}
	&&S[\vb{r}(t)](\vb{r}_b,\vb{r}_a,t_b,t_a)\\
	&&=S^{(0)}[\vb{r}(t)](\vb{r}_b,\vb{r}_a,t_b,t_a)-\int_{t_a}^{t_b}\,\Delta V(\vb{r}(t),t)\,dt\,.\nonumber
	\label{eq:30}
\end{eqnarray}
$S[\vb{r}(t)]$ is the classical action functional defined for the path $\vb{r}(t)$, connecting the starting point $\vb{r}_a=\vb{r}(t_a)$ to the end point
$\vb{r}_b=\vb{r}(t_b)$. In order to establish gauge invariance and to ensure operator ordering, we proceed with the discretization
of the time variable. The time interval $T=t_b-t_a$ is divided into $N$ small discrete intervals. The time lattice is divided into $N+1$ equidistant points denoted
by $u_j$, as in Fig.~\ref{fig:timelat}, such that 
\begin{align*}
	u_j&=t_a+jt_j\,,\quad \text{where,}\,\quad t_j=\frac{t_b-t_a}{N}=u_j-u_{j-1}\,,\\
	u&=\sum t_j\,.
\end{align*} 
With this discretization of the time interval, the lattice action reads
\begin{align}
	&S^{(N)}[\vb{r}_1,\dots,\vb{r}_{N-1}](\vb{r}_b,\vb{r}_a,t_b,t_a)\nonumber\\
	&=S^{(0;N)}[\vb{r}_1,\dots,\vb{r}_{N-1}](\vb{r}_b,\vb{r}_a,t_b,t_a)\nonumber\\
	&\qquad\qquad\qquad\qquad+\Delta S^{(N)}[\vb{r}_1,\dots,\vb{r}_{N-1}](\vb{r}_b,\vb{r}_a,t_b,t_a)\,,\nonumber\\
	&=S^{(0;N)}[\vb{r}_1,\dots,\vb{r}_{N-1}](\vb{r}_b,\vb{r}_a,t_b,t_a)-t_j\sum_{j=0}^{N-1}\,\Delta V(\vb{r}_j,u_j)\,,\label{eq:31}
\end{align}
where $S^{(0;N)}$ is the unperturbed action in the time-lattice definition.
Using this time-lattice version of the action, if we construct the action integral for the action perturbed by an arbitrary potential, we obtain
\begin{eqnarray}
	e^{\frac{i}{\hbar}S^{(N)}}=e^{\frac{i}{\hbar}S^{(0;N)}}+e^{\frac{i}{\hbar}\Delta S^{(N)}}\,.\label{eq:32}
\end{eqnarray}
We expand the perturbation exponential up to all orders as
\begin{eqnarray}
	&&e^{\frac{i}{\hbar}\Delta S^{(N)}}\nonumber\\
	&&=\sum_{n=0}^{\infty}\frac{t_j^{n}}{n!}\left[\sum_{j=0}^{N-1}\frac{\Delta V(\vb{r}_j,u_j)}{i\hbar}\right]^{n}\nonumber\\
	&&=\sum_{n=0}^{\infty}\frac{t_j^{n}}{n!}\left[\sum_{j_n=0}^{N-1}\dots\sum_{j_1=0}^{N-1}\frac{\Delta V(\vb{r}_{j_n},u_{j_n})}{i\hbar}\dots\frac{\Delta V(\vb{r}_{j_1},u_{j_1})}{i\hbar}\right]\nonumber\\
	&&=\sum_{n=0}^{\infty}t_j^{n}\left[\sum_{j_n=0}^{N-1}\dots\sum_{j_1=0}^{N-1}\frac{\Delta V(\vb{r}_{j_n},u_{j_n})}{i\hbar}\dots\frac{\Delta V(\vb{r}_{j_1},u_{j_1})}{i\hbar}\right]\nonumber\\
	&&\quad\qquad\times\left[1+\mathcal{O}\left(\frac{1}{N}\right)\right]\,,\label{eq:33}
\end{eqnarray}
where the correction terms of order ${O}\left(\frac{1}{N}\right)$ are to take care of the repeated indices.
In the limit $N\rightarrow\infty$, Eq.~(\ref{eq:33}) becomes
\begin{eqnarray}
	&&\sum_{n=0}^{\infty}t_j^{n}\left[\sum_{j_n=0}^{N-1}\dots\sum_{j_1=0}^{N-1}\frac{\Delta V(\vb{r}_{j_n},u_{j_n})}{i\hbar}\dots\frac{\Delta V(\vb{r}_{j_1},u_{j_1})}{i\hbar}\right]\nonumber\\
	&&\qquad\quad\times\left[1+\mathcal{O}\left(\frac{1}{N}\right)\right]\nonumber\\
	&&{\approx}_{N\rightarrow\infty}\int_{t_a}^{t_b}du_n\int_{t_a}^{u_n}du_{n-1}\dots\nonumber\\
	&&\qquad\qquad\times\int_{t_a}^{u_2}du_{1}\frac{\Delta V(\vb{r}_{n},u_{n})}{i\hbar}\dots\frac{\Delta V(\vb{r}_{1},u_{1})}{i\hbar}\,.\label{eq:34}
\end{eqnarray}

In order to simplify the notation in Eq.~(\ref{eq:34}), in the context of perturbation theory, we redefine the time intervals as
\begin{align*}
	u_{(0)}=u_0=t_a\,,\qquad u_{(n+1)}=u_{N}=t_b\,,\nonumber\\
	u_{(k)}=u_{j_k}\,,\quad	\vb{r}_{(k)}=\vb{r}_{j_k}\,,\qquad k=1,\dots, n\,.
\end{align*}
At each order of perturbation, the time interval $T=t_b-t_a$ is divided into $n+1$ subintervals. The subscript $j$ defines the discretization of the
time lattice into $N$ discrete intervals, and the subscript $k$ denotes the order of perturbation. Thus, we discretize time, and calculate the path integrals,
for each order of perturbation theory. As such, the lattice unperturbed action at each order of perturbation theory is given as
\begin{eqnarray}
	&&S^{(0;N)}[\vb{r}_1,\dots,\vb{r}_{N-1}](\vb{r}_b,\vb{r}_a,t_b,t_a)\label{eq:35}\\
	&&\qquad=\sum_{k=0}^{n}	S^{(0;N)}[\vb{r}_1,\dots,\vb{r}_{N-1}](\vb{r}_{(k+1)},\vb{r}_{(k)};u_{(k+1)},u_{(k)})\,.\nonumber
\end{eqnarray}
We can define the time-lattice division as per the following equation:
\begin{eqnarray}
	&&T^{N}(t_b,t_a)=\label{eq:36}\\
	&&\qquad\{T^{(N_0)}(t_a,u_{(k)});T^{(N_1)}(u_1,u_{2});\dots;T^{(N_n)}(u_n,t_{b})\}\,,\nonumber
\end{eqnarray}
where 
\begin{eqnarray}
	&&T^{(N_k)}(u_{(k+1)},u_{(k)})\label{eq:37}\\
	&&=(u_{k,j_0}\equiv u_{(k)},u_{k,j_1},\dots,u_{k,j_{N_{k-1}}}u_{k,j_{N_{k}}}\equiv u_{(k+1)})\,.\nonumber
\end{eqnarray}
the entire interval has been divided into $n+1$ subintervals and each of these subintervals, $[u_{(k+1)},u_{(k)}]$ have been discretized into the usual
$N_k$ parts, at each order of perturbation theory.

\begin{figure}
	\centering
	\includegraphics[width=0.45\textwidth]{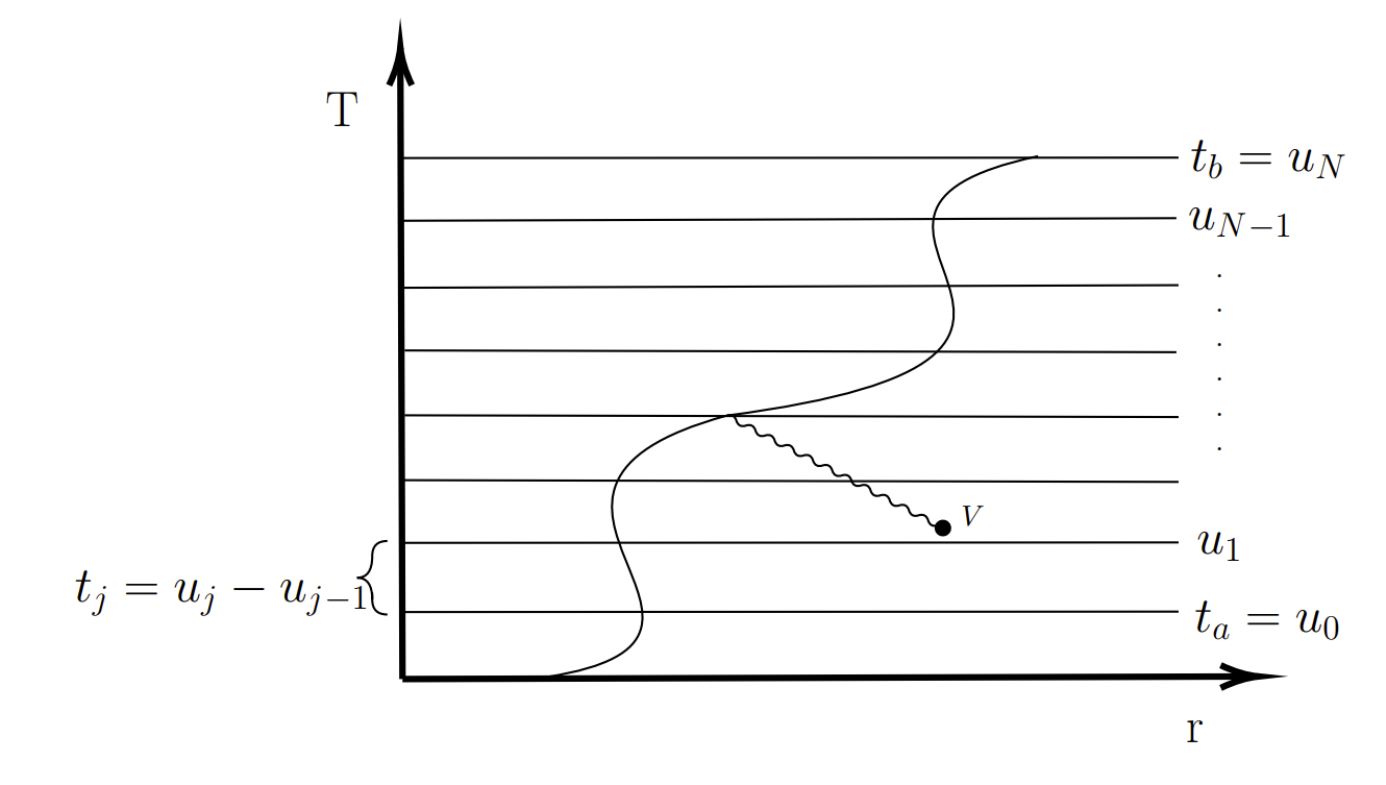}
	\caption{Time-lattice division for first-order perturbation.}
	\label{fig:pertime}
\end{figure}

Owing to this time-lattice divisions, the integration measure for the path integrals can be given in terms of a partial measure for each order of perturbation
\begin{equation}
	\int_{\vb{r}_a}^{\vb{r}_b}\mathcal{D}^{N}\vb{r(u)}=\prod_{k=1}^{n}\left[\int d\vb{r}_k\right]\prod_{i=0}^{n}\left[\int_{\vb{r}_{i}}^{\vb{r}_{i+1}}\mathcal{D}^{(N_i)}\vb{r}(u) \right]\,.\label{eq:38}
\end{equation} 
Knowing that the Feynman path integral kernel has the form
\begin{eqnarray}
	&&K(\vb{r}_b,\vb{r}_a;t_b,t_a)\label{eq:39}\\
	&&=\lim_{N\rightarrow\infty}\int_{\vb{r}_a}^{\vb{r}_b}\mathcal{D}\vb{r}(t)\exp\left\{\frac{i}{\hbar}S[\vb{r}(t)](\vb{r}_b,\vb{r}_a,t_b,t_a)\right\}\,,\nonumber
\end{eqnarray}
with the integration measure of Eq.~(\ref{eq:38}) and with the help of Eq.~(\ref{eq:31}) and Eq.~(\ref{eq:34}), the Feynman kernel can be rewritten as
\begin{eqnarray}
	&&K^{(n)}(\vb{r}_b,\vb{r}_a;t_b,t_a)=\int_{t_a}^{t_b}du_n\int_{t_a}^{u_n}du_{n-1}\dots\int_{t_a}^{u_2}du_{1}\nonumber\\
	&&\qquad\qquad\qquad\times\prod_{k=1}^{n}\left[\int d\vb{r}_k\frac{\Delta V(\vb{r}_k,u_k)}{i\hbar}\right]\nonumber\\
	&&\qquad\qquad\qquad\qquad\times\prod_{i=0}^{n}\left[K^{(0)}(\vb{r}_{i+1},\vb{r}_i;u_{i+1},u_i)\right]\,,
	\label{eq:40}
\end{eqnarray}
where the kernel $K^{(0)}$ corresponds to the unperturbed action $S^{(N,0)}[\vb{r}(u)]$, defined in the limit $N\rightarrow\infty$. In case of systems
with a time-independent perturbing potential, the Feynman kernel in Eq.~(\ref{eq:40}) can be expressed as
\begin{eqnarray}
	&&K^{(n)}(\vb{r}_b,\vb{r}_a;T)=\prod_{k=1}^{n}\left[\int d\vb{r}_k\frac{\Delta V(\vb{r}_k)}{i\hbar}\right]	\label{eq:41}\\
	&&\quad\qquad\times\prod_{i=0}^{n}\left[\int_{0}^{\infty}dT_i\,K^{(0)}(\vb{r}_{i+1},\vb{r}_i;T_i)\right]\delta\left(T-\sum_{\gamma=0}^{n}T_{\gamma}\right)\,.\nonumber
\end{eqnarray}
Correspondingly, the $n$-th order energy-dependent Green's function is obtained by taking the Fourier transform of the propagator in the above equation with respect to time, yielding
\begin{eqnarray}
	&&G^{(n)}(\vb{r}_b,\vb{r}_a;E)\label{eq:42}\\
	&&\qquad=\prod_{k=1}^{n}\left[\int_{0}^{\infty}d\vb{r}_k\,\Delta V(\vb{r}_k)\right]\prod_{i=1}^{n}\left[G^{(0)}(\vb{r}_{i+1},\vb{r}_i;E)\right]\,.\nonumber
\end{eqnarray}

\section{Uehling contribution to bound-state energy shifts}
\label{sec:uehling}
The infinite summation of a perturbation expansion using path integrals developed in the last Section is now implemented to study the correction to the binding energy of an electron
due to the perturbing Uehling potential. As in the Coulomb case, we start from the Dirac equation with a given potential $V(\vb{r})$. The time-independent Dirac equation
is customarily expressed, in natural units ($c=1$, $\hbar=1$), as
\begin{equation}
	(E-\boldsymbol{\alpha} \cdot \hat{\boldsymbol{p}}-\beta m-V(\vb{r})) \Psi=0 \,,
	\label{eq:43}
\end{equation}
where $\Psi$ is the bispinor wave function of the Dirac particle, $m$ is the mass, $\boldsymbol{\alpha}$ and $\beta$ are the usual $ 4\cross4$ Dirac matrices, and $E$ is the energy.

In order to calculate the contribution of the Uehling correction to the binding energy, we treat the Uehling potential defined in Eq.~(\ref{eq:29}) as the perturbation.
As in Eq.~(\ref{eq:31}), the lattice-perturbed action is given as a sum of the unperturbed action and the action due to the perturbing potential.
For our case, i.e., in the Furry picture, we consider the action due to the nuclear Coulomb potential as the unperturbed action
$S^{(0)}$~\cite{INOMATA1982387,PhysRevLett.53.107}. The unperturbed action of the bound electron is
\begin{equation}
	S^{(0)}(t_j)=\frac{m(\Delta r_j)^2}{2t_j}-\frac{\lambda(\lambda+1)t_j}{2mr_jr_{j-1}}+\frac{Ze^2Et_j}{mr_j}+\frac{(E^2-m^2)t_j}{2m}\,,
	\label{eq:44}
\end{equation}
where, $\lambda\equiv|\gamma|+\frac{1}{2}(\text{sign}\,\gamma -1)$, $\gamma$ being the eigenvalue of the Martin-Glauber operator~\cite{Martin1958},
which is defined as $\mathcal{L}=-(\beta K+iZe^2\alpha_r)$, with $K$ being the Dirac operator and $\alpha_r=\boldsymbol{\alpha}\cdot\vb{r}/r$~\cite{PhysRevLett.53.107}.

Coupled with the perturbing potential this yields the lattice-perturbed action. From Eq.~(\ref{eq:31}) we obtain
\begin{eqnarray}
	&&S[\vb{r}(t)](\vb{r}_b,\vb{r}_a,t_b,t_a)=\frac{m(\Delta r_j)^2}{2t_j}-\frac{\lambda(\lambda+1)t_j}{2mr_jr_{j-1}}+\frac{Ze^2Et_j}{mr_j}\nonumber\\
	&&\qquad\qquad\qquad+\frac{(E^2-m^2)t_j}{2m}-\int_{t_a}^{t_b}\,\Delta V(\vb{r}(t))\,dt\,,
	\label{eq:45}
\end{eqnarray}
where 
\begin{eqnarray}
	&&\Delta V(\vb{r})=U_{\rm VP}^{\rm Ueh}
	\label{eq:46}
\end{eqnarray}
is the perturbing Uehling potential.

The corresponding Feynman kernel can be derived using Eq.~(\ref{eq:41}) and eventually the energy-dependent Green's function is given by Eq.~(\ref{eq:42}). Thus,
the $n$-th order Green's function is obtained in terms of the Dirac-Coulomb propagator, which has been determined using the path integral formalism~\cite{PhysRevLett.53.107}, as
\begin{eqnarray}
	&&G^{(n)}(\vb{r}_b,\vb{r}_a;E)\nonumber\\
	&&=\prod_{k=1}^{n}\left[\int_{0}^{\infty}d\vb{r}_k\,\Delta V(\vb{r}_k)\right]\prod_{i=0}^{n}\left[G^{(0)}(\vb{r}_{i+1},\vb{r}_i;E)\right]\nonumber\\
	&&=\prod_{k=1}^{n}\Biggl[\int_{0}^{\infty}d\vb{r}_k\,\Biggl\{-\frac{2\alpha Ze}{3\pi r_k}\Biggl[\left(1+\frac{r_k^2}{3\alpha^2}\right)K_0   -\frac{r_k}{6\alpha}Ki_1\nonumber\\
	&&\qquad\qquad-\left(\frac{5}{6}+\frac{r_k^2}{3\alpha^2}\right)Ki_2\Biggr]\Biggr\}\Biggr]\nonumber\\
	&&\times\prod_{i=0}^{n}\Biggl[\sum_{j,\kappa}\frac{\Gamma(p+\lambda+1)}{2\iota kr_1r_2\Gamma(2\lambda+2)}W_{-p,\lambda+1/2}(-2\iota kr_1)\nonumber\\
	&&\times\Biggl\{\left[m-\frac{\kappa E}{\gamma}\right]M_{-p,\lambda+1/2}(-2\iota kr_2)\Omega_{\kappa,\kappa}^j(\theta_2\phi_2|\theta_1\phi_1)\beta^2\nonumber\\
	&&-k\,\tilde{\gamma} M_{-p,\bar{\lambda}+1/2}(-2\iota kr_2)\Omega_{\kappa,-\kappa}^j(\theta_2\phi_2|\theta_1\phi_1)\beta^2\gamma_5
	\Biggr\}\Biggr]\,.
	\label{eq:47}
\end{eqnarray}
The energy-dependent Green's function is then 
\begin{eqnarray}
	G(\vb{r}_b,\vb{r}_a;E)=\sum_{n=0}^{\infty}G^{(n)}(\vb{r}_b,\vb{r}_a;E)\,.
	\label{eq:48}
\end{eqnarray}

Considering up to second order, the Green's function in Eq.~(\ref{eq:48}) can be written as
\begin{eqnarray}
	G(\vb{r}_b,\vb{r}_a;E)&&=G^{(0)}(\vb{r}_b,\vb{r}_a;E)\nonumber\\
	&&+G^{(1)}(\vb{r}_b,\vb{r}_a;E)+G^{(2)}(\vb{r}_b,\vb{r}_a;E)\,,
	\label{eq:49}
\end{eqnarray}
where
\begin{eqnarray}
	&&G^{(1)}(\vb{r}_b,\vb{r}_a;E)=\left[\int_{0}^{\infty}d\vb{r}_1\,\Delta V(\vb{r}_1)\right]\prod_{i=0}^{1}\left[G^{(0)}(\vb{r}_{i+1},\vb{r}_i;E)\right]\,,\nonumber\\
		\label{eq:50}
\end{eqnarray}
and
\begin{eqnarray}
	&&G^{(2)}(\vb{r}_b,\vb{r}_a;E)=\prod_{k=1}^{2}\left[\int_{0}^{\infty}d\vb{r}_k\,\Delta V(\vb{r}_k)\right]\nonumber\\
	&&\qquad\qquad\qquad\times\prod_{i=0}^{2}\left[G^{(0)}(\vb{r}_{i+1},\vb{r}_i;E)\right]\,.\nonumber\\
	\label{eq:51}
\end{eqnarray}

In analogy to the H-atom problem~\cite{PhysRevLett.53.107}, the discrete energy spectrum can be obtained from the poles of the spectral function,
which is defined as
\begin{eqnarray}
	&&G(E)=\int G(\vb{r},\vb{r};E)  \,d\vb{r} \label{eq:52}
\end{eqnarray}
with the Green's function [e.g. in Eq.~{\ref{eq:48}}] in coordinate representation. The form of the Green's function in Eq.~(\ref{eq:47}) does not allow
a direct evaluation, therefore, we apply the basis-state decomposition of the spectral function in terms of the perturbed states $\phi_n$
and the corresponding eigenenergies $E_n$, given as
\begin{eqnarray}
	G(E)=\int\sum_n\frac{\phi_n(\vb{r})\phi^{\dagger}_n(\vb{r})}{E-E_n(1-i0^{+})}\,d\vb{r}\,.
	\label{eq:53}
\end{eqnarray}
In analogy to Refs.~\cite{messiah1961quantum,PhysRevA.16.863,Shabaev1990,Shabaev_2002}, we introduce the spectral function projected to a single reference state $\ket{a}$,
$G_a(E)=\bra{a}G\ket{a}$. This function $G(E)$ possesses a pole at the eigenenergy $E_a$ of the reference state:
\begin{eqnarray}
	G_a(E) \approx \frac{C_a}{E-E_a}\,,
	\label{eq:54}
\end{eqnarray}
where the constant $C_a$ is the residue term. The energies are now determined using complex contour integration by considering a small contour $\Gamma$ which
encircles an isolated pole at the perturbed bound-state energy $E_a$. From the formulas
\begin{eqnarray}
	&&\frac{1}{2\pi i}\oint_{\Gamma} dE\,E\,G_a(E)= E_aC_a
	\label{eq:55}\,,\\\nonumber\\
	&&	\text{and}\nonumber\\\nonumber\\
	&&	\frac{1}{2\pi i}\oint_{\Gamma} dE\,G_a(E)= C_a \,.
	\label{eq:56}
\end{eqnarray}
one can extract the level energy as
\begin{eqnarray}
	E_a=\frac{\frac{1}{2\pi i}\oint_{\Gamma} dE\,E\,G_a(E)}{\frac{1}{2\pi i}\oint_{\Gamma} dE\,G_a(E)}\,.
	\label{eq:57}
\end{eqnarray}
It is more practical to directly adopt the expression for the perturbative energy shift 
\begin{eqnarray}
	\Delta E_a=\frac{\frac{1}{2\pi i}\oint_{\Gamma} dE\,\Delta E\,\Delta G_a(E)}{1+\frac{1}{2\pi i}\oint_{\Gamma} dE\,\Delta G_a(E)}\,,
	\label{eq:58}
\end{eqnarray}
where
\begin{eqnarray}
\Delta E_a=	\Delta E_a^{(1)}+\Delta E_a^{(2)}+\dots \,,\nonumber\\
\Delta G_a=	\Delta G_a^{(1)}+\Delta G_a^{(2)}+\dots \,,\nonumber
\end{eqnarray}
are expansions in terms of the perturbing potential. We expand  Eq.~(\ref{eq:58}) in a geometric series and obtain, to first order,
\begin{eqnarray}
	\Delta E_a^{(1)}=\frac{1}{2\pi i}\oint_{\Gamma} dE\,\Delta E\,\Delta G_a^{(1)}(E)\,.
	\label{eq:59}
\end{eqnarray}
Thus the perturbative change of the spectral function is 
\begin{eqnarray}
	\Delta G_a^{(1)}(E)\sim\frac{\bra{a}V\ket{a}}{(E-E_a)^2}\,,
	\label{eq:60}
\end{eqnarray} 
and from Eq.~(\ref{eq:59}), one obtains the perturbative energy shift
\begin{eqnarray}
	\Delta E_a^{(1)}=\bra{a}V\ket{a}\,,
	\label{eq:61}
\end{eqnarray}
with $\Delta E_a= E-E_a^{(0)}$. The second-order energy shift can be obtained from the next terms of the geometric expansion of Eq.~(\ref{eq:58}) as
\begin{eqnarray}
	&&\Delta E_a^{(2)}=\frac{1}{2\pi i}\oint_{\Gamma} dE\,\Delta E\,\Delta G_a^{(2)}(E)\label{eq:63}\\
	&&-\left(\frac{1}{2\pi i}\oint_{\Gamma} dE\,\Delta E\,\Delta G_a^{(1)}(E)\right)\left(\frac{1}{2\pi i}\oint_{\Gamma} dE\,\Delta G_a^{(1)}(E)\right)\,.\nonumber
\end{eqnarray}
The first term in Eq.~(\ref{eq:63}) is the irreducible (non-degenerate) part and the second term is the reducible (degenerate) part.
For the irreducible part, we obtain
\begin{eqnarray}
	&&\frac{1}{2\pi i}\oint_{\Gamma} dE\,\Delta E\,\Delta G_a^{(2)}(E)=\nonumber\\
	&&\qquad\qquad\qquad\sum_{i\neq a}\frac{\bra{a}U_{\rm VP}^{\rm Ueh}\ket{i}\bra{i}U_{\rm VP}^{\rm Ueh}\ket{a}}{E_a-E_i}\,,
	\label{eq:64}
\end{eqnarray}
and the reducible part is given as
\begin{eqnarray}
	&&\left(\frac{1}{2\pi i}\oint_{\Gamma} dE\,\Delta E\,\Delta G_a^{(1)}(E)\right)\left(\frac{1}{2\pi i}\oint_{\Gamma} dE\,\Delta G_a^{(1)}(E)\right)\nonumber\\
	&&\qquad=\Delta E_a^{(1)}\left(\frac{1}{2\pi i}\oint_{\Gamma} dE\,\frac{\bra{a}U_{\rm VP}^{\rm Ueh}\ket{a}}{(E-E_a)^2}\right)\nonumber\\
	&&\qquad=\bra{a}U_{\rm VP}^{\rm Ueh}\ket{a}\bra{a}(dU_{\rm VP}^{\rm Ueh}/dE)_{E=E_a}\ket{a}\,.
	\label{eq:65}
\end{eqnarray}
The Uehling potential, as defined in Eq.~(\ref{eq:46}), is independent of the bound-state energy and as such the derivative term in Eq.~(\ref{eq:65}) vanishes,
and the second-order correction is then given by the usual Rayleigh-Schr\"odinger perturbative expression
\begin{eqnarray}
	\Delta E_a^{(2)}=\sum_{i\neq a}\frac{\bra{a}U_{\rm VP}^{\rm Ueh}\ket{i}\bra{i}U_{\rm VP}^{\rm Ueh}\ket{a}}{E_a-E_i}\,,
	\label{eq:66}
\end{eqnarray}
in agreement with Ref.~\cite{Yerokhin2008}.

\section{Numerical results}
\label{sec:results}

To confirm the validity of our approach, in Table~\ref{tab:results}, we present numerical results of first-order perturbative Uehling shifts [Eq.~(\ref{eq:61})] for heavy
hydrogenlike ions and certain alkali ions, and compare them earlier evaluations of this correction (see e.g. Ref.~\cite{Yerokhin2015}). A good agreement can be stated. We assume
for simplicity a point-like nuclear charge distribution.

The results in Table~\ref{tab:results} reiterate the well-known fact that with the increase in the atomic number, the VP effects become more
and more pronounced; VP effects scale as $\sim \left( Z \alpha \right)^4$ to leading order. For most elements and charge states, the VP shift is experimentally discernible
with modern Penning-trap mass spectrometric methods with experimental uncertainties on the 1-eV level or below~\cite{Rischka2020,Schuessler2020,Filianin2021,Kromer2022}.
We thus identify ions which allow, for the first time, a test of QED via measuring the electronic binding energy of a valence electron by determining the mass
difference of two ions, one with a single valence electron and one without.

\begin{table*}
\begin{ruledtabular}
\begin{tabular}{rlllllr}
\label{tab:results}
$Z$ & \multicolumn{5}{c}{Uehling correction [eV]}\\
\cline{2-6}
    &   $1s$    & $2s$      & $3s$    & $2p_{1/2}$ & $2p_{3/2}$    & Ref. \\
    &  H-like   & Li-like   & Na-like & B-like     & N-like        &   \\
\hline
54  &  -7.3516  &  -1.0272  &  -0.306 &  -0.04931  &   -0.00555    &   \\
    &  -7.35163 &  -1.02717 &         &  -0.049313 &   -0.005545   &   \cite{Yerokhin2015}\\
82  & -51.875   &  -8.4760  &  -2.466 &  -1.0938   &   -0.063801   &   \\
    & -51.8751  &  -8.47600 &         &  -1.09378  &   -0.0638007  &   \cite{Yerokhin2015}\\
92  & -97.996   & -17.309   &  -4.878 &  -2.9954   &   -0.12653    &   \\
    & -97.9960  & -17.3092  &         &  -2.99533  &   -0.126532   &   \cite{Yerokhin2015}
\end{tabular}
\caption{Uehling vacuum polarization corrections for certain H-like heavy ions, in units of eV.}
\end{ruledtabular}
\end{table*}

The values in the Table also show that it is not only H-like ions in the $1s$ ground state which
feature VP effects well observable by these experimental methods, but also excited states possess
observable radiative shifts. E.g. the VP shift given for the $2s$ state approximates the VP correction to the (negative of the) binding energy of a Li-like ion, which
can be spectrometrically determined by measuring the mass difference of the Li- and
the He-like ions in their ground states. In our approach presented here, we fully neglect many-electron (screening) effects, which, for the high-$Z$ ions
presented here, is a justified first approximation, as they contribute with a relative order of $\frac{1}{Z}$. Our path
integral formalism however may be extended in future to many-electron systems, by allowing the exchange of photons between different electrons.

Similarly, the VP results given for the $3s$ state approximate the radiative shift of the binding energy of the
valence electron in the Na-like charge state, and the values given for the $2p_{1/2}$ and $2p_{3/2}$ orbitals give
the first approximation for B- and N-like ions, respectively. Ions in these charge states are easier to produce experimentally,
and Na- and B-like very heavy species feature observable VP contributions, enabling a proof-of-the-principle demonstration of QED
tests via mass measurements.

\section{Summary}

We have calculated the VP correction to energy levels of highly charged ions employing a path integral formalism.
First, the evaluation of the leptonic loop correction to the photon propagator by means of the Dyson-Schwinger equation is summarized.
The effective potential describing the correction to the nuclear potential in the lowest order in $Z\alpha$, i.e. the Uehling potential, is used in an analytical form given
by Frolov and Wardlow~\cite{Frolov2012}.
The contribution to the energy level of a hydrogenlike ion due to the perturbing Uehling potential is derived from the perturbed action using a relativistic path integral method.
We show that the VP level shifts -- or any energy shift induced by a local potential -- can be extracted from the poles of the Green's function. The VP correction is given
numerically for a range of heavy ions, concluding that in sufficiently highly charged ions, the VP effect is observable with state-of-the-art mass spectrometric methods.

\section*{Acknowledgements}

Supported by the Deutsche Forschungsgemeinschaft (DFG, German Research Foundation) – Project-ID 273811115 – SFB 1225.

\end{document}